\documentclass[
twocolumn,
longbibliography,
superscriptaddress,
amssymb,
amsmath
]{revtex4-2}

\usepackage[dvips]{graphicx}
\usepackage{epsf}
\usepackage{dcolumn}
\usepackage{bm}
\usepackage{bm, color}
\usepackage{bbm}
\usepackage{lineno}
\usepackage{braket}
\usepackage{amsbsy}
\usepackage{xr}
\usepackage{easyReview}

\usepackage[
colorlinks=true,
linkcolor=magenta,
urlcolor=magenta,
citecolor=magenta,
filecolor=blue,
pagecolor=blue,
linktocpage=true,
]{hyperref}

\begin{document}

	\title{Prediction of dual quantum spin Hall insulator in NbIrTe$_4$ monolayer}
	
	\author{Xiangyang Liu}
	\affiliation{School of Materials Science and Engineering, University of Science and Technology of China, Shenyang 110016, China.}
	\affiliation{Shenyang National Laboratory for Materials Science, Institute of Metal Research, Chinese Academy of Sciences, Shenyang 110016, China.}

	\author{Junwen Lai}
	\affiliation{School of Materials Science and Engineering, University of Science and Technology of China, Shenyang 110016, China.}
	\affiliation{Shenyang National Laboratory for Materials Science, Institute of Metal Research, Chinese Academy of Sciences, Shenyang 110016, China.}
	
	\author{Jie Zhan}
	\affiliation{School of Materials Science and Engineering, University of Science and Technology of China, Shenyang 110016, China.}
	\affiliation{Shenyang National Laboratory for Materials Science, Institute of Metal Research, Chinese Academy of Sciences, Shenyang 110016, China.}
	
	\author{Tianye Yu}
	\affiliation{Shenyang National Laboratory for Materials Science, Institute of Metal Research, Chinese Academy of Sciences, Shenyang 110016, China.}
	
\author{Wujun Shi}
\affiliation{Center for Transformative Science, ShanghaiTech University, Shanghai 201210, China}
\affiliation{Shanghai High Repetition Rate XFEL and Extreme Light Facility (SHINE), ShanghaiTech University, Shanghai 201210, China}
	
	\author{Peitao Liu}
	\affiliation{School of Materials Science and Engineering, University of Science and Technology of China, Shenyang 110016, China.}
	\affiliation{Shenyang National Laboratory for Materials Science, Institute of Metal Research, Chinese Academy of Sciences, Shenyang 110016, China.}
	
	\author{Xing-Qiu Chen}
	\email{xingqiu.chen@imr.ac.cn}
	\affiliation{School of Materials Science and Engineering, University of Science and Technology of China, Shenyang 110016, China.}
	\affiliation{Shenyang National Laboratory for Materials Science, Institute of Metal Research, Chinese Academy of Sciences, Shenyang 110016, China.}
	
	\author{Yan Sun}
	\email{sunyan@imr.ac.cn}
	\affiliation{School of Materials Science and Engineering, University of Science and Technology of China, Shenyang 110016, China.}
	\affiliation{Shenyang National Laboratory for Materials Science, Institute of Metal Research, Chinese Academy of Sciences, Shenyang 110016, China.}

	\begin{abstract} 
	 Dual quantum spin Hall insulator (QSHI) is a newly discovered topological state
	 in the 2D material TaIrTe$_4$, exhibiting both a traditional $Z_2$ band gap at charge neutrality point and a van Hove singularity (VHS) induced correlated $Z_2$ band gap with weak doping.
     Inspired by the recent progress in theoretical understanding and experimental
	 measurements, we predicted a promising dual QSHI in the counterpart material of the 
	 NbIrTe$_4$ monolayer by first-principles calculations. In addition to the well-known band inversion at the charge neutrality point, two new band inversions were
	 found after CDW phase transition when the chemical potential is near the VHS, one direct and one indirect $Z_2$ band gap.  
	 The VHS-induced non-trivial band gap is around 10 meV, much larger than that from TaIrTe$_4$. Furthermore, since the new generated band gap is mainly dominated
	 by the $4d$ orbitals of Nb, electronic correlation effects should be relatively stronger in NbIrTe$_4$ as compared to TaIrTe$_4$. Therefore, the dual QSHI state in the NbIrTe$_4$ 
	 monolayer is expected to be a good platform for investigating the interplay between topology
	 and correlation effects.

	\end{abstract}
	
	\maketitle
	\section{Introduction and motivation}
    Two-dimensional (2D) materials with VHS near the
    Fermi level are ideal platforms for studying the interplay between topological
    electronic structure and correlation effects\cite{wang2020correlated,liu2015electronic,hsu2017topological,wei2024,kim2018,xu2021,wu2021,lee2021,kang2020,xu2013,sun2022}. 
    As a typical layered material, TaIrTe$_4$ monolayer was predicted to be
    a 2D topological insulator with a band gap around 32 meV\cite{guo2020quantum,liu2017van,marrazzo2019relative}. 
    Very recently, a new quantum topological state of duel QSHI was discovered in the TaIrTe$_4$ monolayer by the observation of quantized longitudinal conductance, which is rare in other QSHIs\cite{tang2024dual}. It was found that, in addition to the QSHI state at the charge neutrality point, two new non-trivial Z$_2$ gaps are generated when the Fermi level shifts
    to the VHS points by weak electron and hole doping effects, respectively.

    Owing to the divergence of density of states (DOSs), a CDW phase transition is induced when the Fermi level shifts to the VHSs, resulting in the correlated $Z_2$ band gap. Different from traditional $Z_2$ band gaps, the VHSs generated topological band gap is expected to naturally relate to the novel	correlated electronic topological states of the fractional quantum spin Hall effect and helical quantum spin liquid\cite{maciejko2010fractional, santos2011time, goerbig2012fractional, li2014fractional,wang2016time,barkeshli2019symmetry, park2010dirac}. Though the VHSs generated $Z_2$ band gap 
	and integer quantized longitudinal conductance can exist in TaIrTe$_4$, the corresponding fractional topology and other strongly correlated topological states have, so far, not been observed. Therefore, it is necessary to find out some more materials candidates that host both topology and correlation\cite{ghimire2020topology,paschen2021quantum,hasan2010a,teng2022,yin2022,yin2020,yin2018,kang2020a,guo2024}. One possible reason for the absence of strongly correlated topological phenomena in TaIrTe$_4$ is the weak strength of electron correlation.
    With this inspiration, we analyzed the dual $Z_2$ topological state in NbIrTe$_4$, an isoelectronic counterpart of TaIrTe$_4$. Due to the difference between Ta-5d and Nb-4d orbitals near VHSs, the electronic correlation effect is expected to be more pronounced in NbIrTe$_4$ than in TaIrTe$_4$. This makes the NbIrTe$_4$ to be a good candidate material for observing the correlated topological phenomena that are absent in TaIrTe$_4$.

	\section{Result and Discussion}
    The crystal structure of a NbIrTe$_4$ monolayer shares the same structural prototype as 1T$^\prime$-MoTe$_2$ monolayer\cite{empante2017chemical} and features a sandwich-like structure, as depicted in Fig. 1(a). In this arrangement, 
    the Nb and Ir atoms are positioned in the middle layer, forming parallel zigzag chains along the $b$-axis. 
    Each Nb or Ir atom is surrounded by six Te atoms, forming a distorted octahedron. 
     Fig. 1(c) shows the energy dispersion along the high-symmetry lines (Fig. 1(b)) calculated by employing Vienna Ab initio Simulation Package (VASP)\cite{hafner2008ab} with generalized gradient approximation (GGA) of the 
     Perdew-Burke-Ernzerhof (PBE) aproximation\cite{garrity2014pseudopotentials}. One can observe that a band inversion exists near the $X$ point with a band gap of around 35 meV, in agreement with previous reports\cite{liu2017van,lv2022}. 
	 In addition, two VHSs are located around $\Gamma$ and $X$, respectively, with the former one above the Fermi level and the latter below the Fermi level. Correspondingly, two narrow peaks of DOSs generated from the VHSs can be found above and below the Fermi level with energy of $\sim$E$_F$ + 60 meV and  $\sim$E$_F$ - 60 meV, respectively. From the orbital projected band structure and 
     DOS, the two VHSs below and above the Fermi level are dominated by the $4d$ orbitals of Nb and hence a strong correlation is expected here. With the above understanding of the electronic band structure near the Fermi level, we projected the Bloch wavefunctions into maximally localized Wannier functions (MLWFs) and constructed effective tight binding model Hamiltonians based on the overlaps of these Wannier orbitals\cite{mostofi2008wannier90}. 

     \begin{figure}[htbp]
    \begin{center}
    \includegraphics[width=0.5\textwidth]{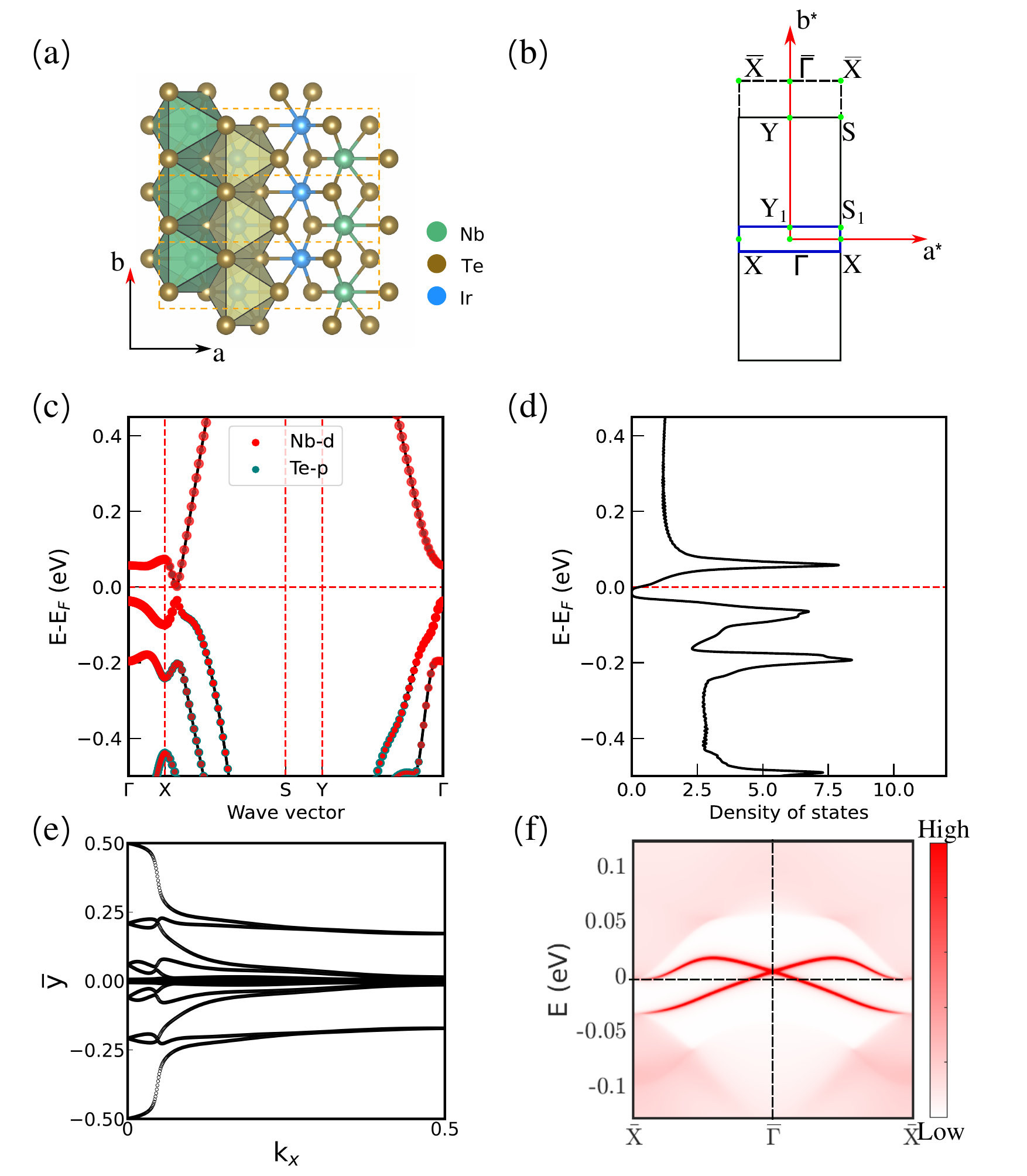}
    \end{center}
    \caption{ Crystal and electronic structure of NbIrTe$_4$ monolayer.
    (a) Top view of the crystal structure.
    (b) Two-dimensional (2D) Brillouin zone (BZ) and its projection along
        the $c$-axis for both the primitive cell and CDW supercell. The high-symmetry lines are denoted as follows: $\Gamma$-X-S-Y for the primitive cell, $\Gamma$-X-S$_1$-Y$_1$ for the 1$\times$ 20 $\times$ 1 supercell, and $\bar{\Gamma}$-$\bar{X}$ for the edge lines along the b-axis.
    (c) Energy dispersion along high-symmetry lines of NbIrTe$_4$ monolayer
        primitive cell.
    (d) Density of states of NbIrTe$_4$ monolayer primitive cell.
    (e) Wannier charge center evolution with Fermi level fixed at
        charge neutrality point.
    (f) Energy dispersion of edge state of NbIrTe$_4$ monolayer
        primitive cell.
     }
    \label{fig1}
    \end{figure}

	 To verify the topological state, we calculated the	 Wannier center evolution and edge state with chemical
	 potential locating at the charge neutrality point\cite{bryant1985surface}. The wanneir center evolution
	 in Fig. 1(e) presents as a zigzag form with changing partner at the edges of 
	 k$_x$=0 and $\pi$/a, which is a typical feature in $Z_2$ topological insulators\cite{soluyanov2011wannier}.
	 Consistent with the bulk state analysis, the edge state with open boundary 
	 condition along the $b$-axis direction shows linear band crossing at the $\Gamma$ point and
	 connects the occupied and unoccupied bulk bands. Hence, the $Z_2$ band gap at
	 the charge neutrality point can be confirmed from both bulk band order and edge Dirac point.

        In order to identify the detailed positions of the VHSs near the charge neutrality point,
	we plot the energy dispersion of band-$n_{occ}+2$ and band-$n_{occ}-2$ 
	in the whole 2D Brillouin zone (BZ), where $n_{occ}$ refers to the index of the highest occupied band in a primitive cell considering the spin. 	As presented in Fig. 2(a), the four	VHS points on band-$n_{occ}+2$ are located at 
	(0.211$\frac{2\pi}{a}$, 0.024$\frac{2\pi}{b}$), 
	and its inversion and mirror partners. Considering the one-dimensional 
	Nb-chains are similar to the Ta-chains in TaIrTe$_4$, the wavevectors along the $b$ 
	direction roughly corresponds to 1$\times$ 20 $\times$ 1 and 1$\times$ 21 $\times$ 1 supercells for the 
	nearby VHSs on band-$n_{occ}+2$ and band-$n_{occ}-2$, respectively. 
 
     \begin{figure}[]
    \begin{center}
    \includegraphics[width=0.5\textwidth]{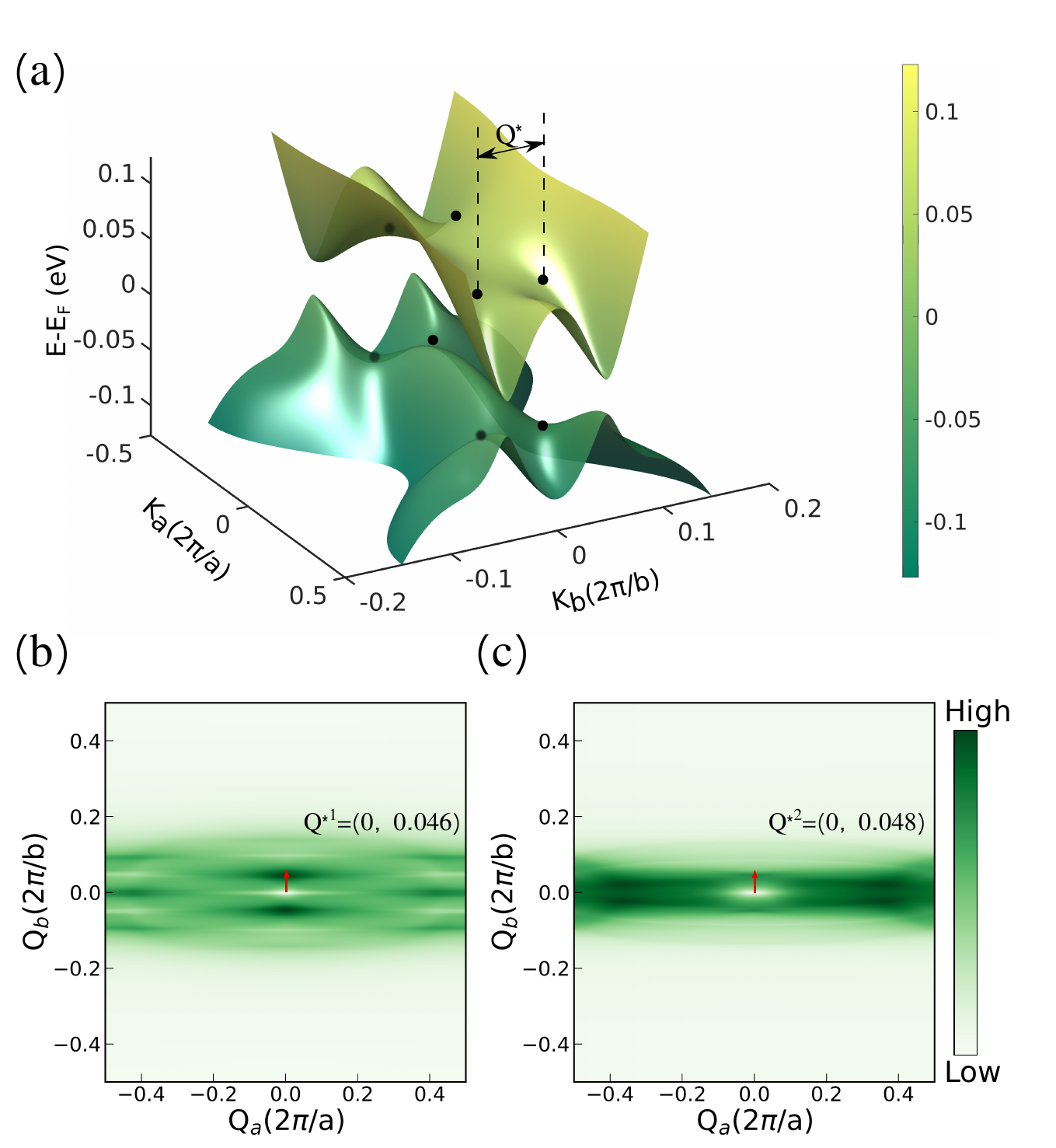}
    \end{center}
    \caption{Band structure and charge susceptibility in NbIrTe$_4$ monolayer 
	 primitive cell.
		 (a) Three dimensional (3D) plot of energy dispersion ($k_{a}-k_{b}-E$) and
		 location of van Hove singularities near the Fermi level.
	 (b, c) Charge susceptibility of  NbIrTe$_4$ with chemical potential
		at VHSs level of the valence and conduction bands, respectively.
    }
    \label{fig2}
    \end{figure}

    We further calculated the charge susceptibility by shifting the chemical potential
	at $\sim$E$_F$ + 60 meV and $\sim$E$_F$ - 60 meV, corresponding to the position of VHSs on band-$n_{occ}+2$ 
	and band-$n_{occ}-2$, respectively. The charge susceptibility $\chi(\textbf{q})$ at a given energy 
	is computed by following the formula of the Lindhard function 
	$\chi(\textbf{q})$ = $\sum\frac{f_ {\textbf{k}}(1-f_{\textbf{k}+\textbf{q}{)}}}{\varepsilon_ {\textbf{k}+\textbf{q}}-\varepsilon_{\textbf{k}}+i\delta}$, where $f$ refers to the Fermi-Dirac distribution, $\varepsilon$ is a momentum-dependent energy and $\delta$ refers to the broadening in data processing. 
 Guided by the fact that the direction of $\textbf{Q}$ is parallel to the Ta-chains in the TaIrTe$_4$ monolayer\cite{tang2024dual}, we specifically chose a vector aligned with the Nb-chains in the NbIrTe$_4$ monolayer prior to further analysis.
 From Fig. 2(b) and (c), one can see that a local maximum of the charge susceptibility is located at the wavevectors of $\textbf{Q}^{*1}=(0,0.046\frac{2\pi}{a})$ and $\textbf{Q}^{*2}=(0,0.048\frac{2\pi}{b})$ for the valence bands and conduction bands, respectively. 
    Therefore, a CDW phase transition may occur when the chemical potential is shifted to near $\sim$E$_F$ + 60 meV and $\sim$E$_F$ - 60 meV via weak doping or gating effects.  

	\begin{figure}[]
    \begin{center}
    \includegraphics[width=0.5\textwidth]{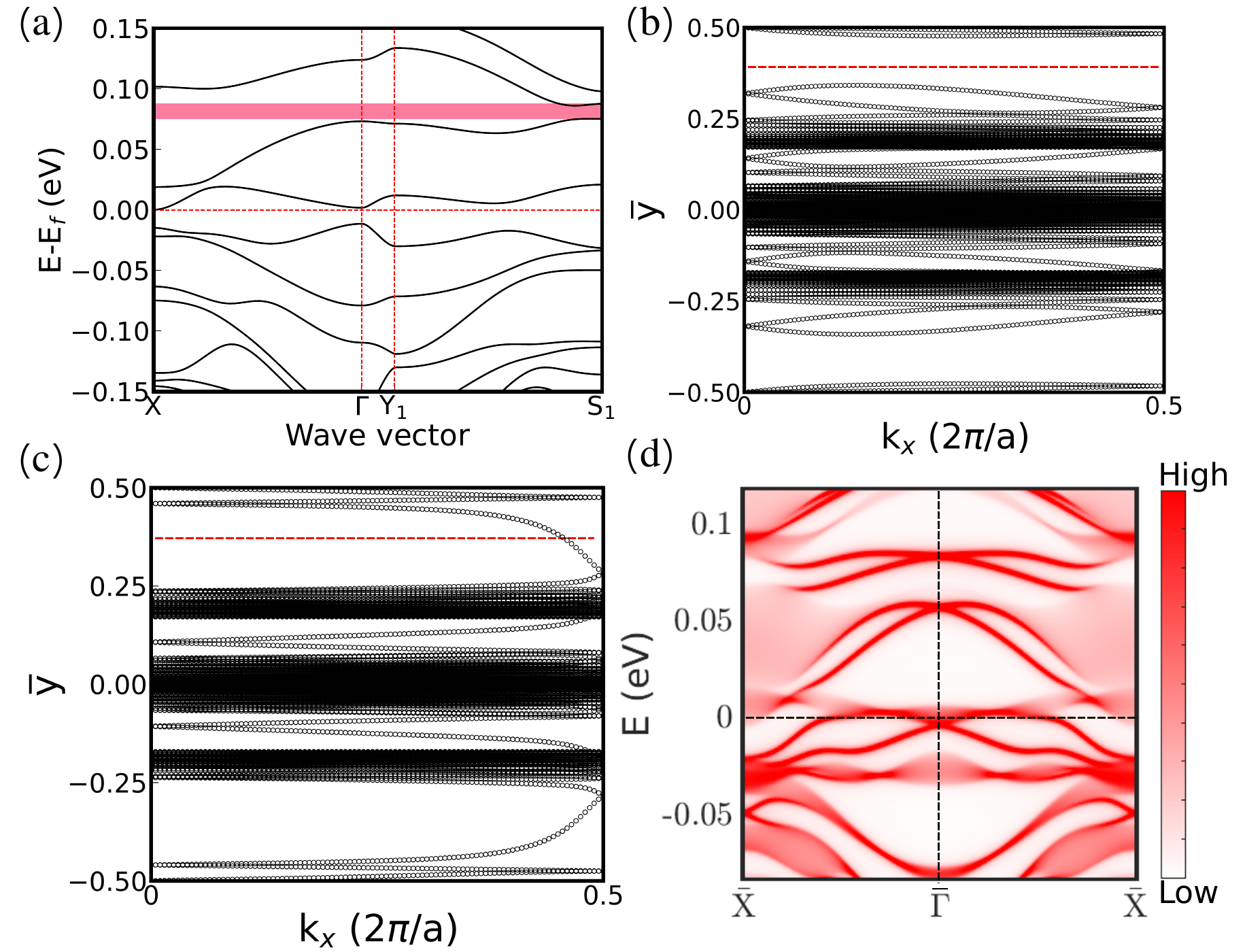}
    \end{center}
    \caption{ Band structure and topological properties in CDW phase.
	(a) Energy dispersion along high symmetry lines. The pink range
	represent the VHS generated $Z_2$ band gap.
	(b-c) Wannier charge center evolutions with Fermi level lying 
	at E$_0$ and E$_0$ + 80 meV, respectively.
	(d) Energy dispersion of edge states.
	}
    \label{fig3}
    \end{figure}

     The wavevectors of $\textbf{Q}^{*1}=(0,0.046\frac{2\pi}{a})$ and $\textbf{Q}^{*2}=(0,0.048\frac{2\pi}{b})$  
	 in Fig. 2 can be described by the supercell of 1$\times$ 21 $\times$ 1 
	 and 1$\times$ 20 $\times$ 1, respectively. To study the 
	 possible CDW phase transition, we constructed an effective  
	 Hamiltonian model for a supercell of 1$\times$ 20 $\times$ 1 based on the overlaps
	 of Wannier orbitals. Following the Fr$\ddot{o}$hlich-Peierls 
	 Hamiltonian approach with supercell along the $b$-axis and only the
	 intra-orbital interaction considered, 
	 the CDW modulated electronic band structure can be described by 
$H=H_{r}+Vcos(Qy)\psi_{r}^{\dagger}\psi_{r}$\cite{frohlich1954theory}. Here, $V$ represents the potential induced by the superlattice. The periodicity of this potential is determined by the parameter $Q$, which can be identified by examining Q$^{*}$ in Fig. 2.

     By increasing the potential $V$, a new band gap emerges near E$_F$ + 80 meV, originating from the original VHSs, as illustrated by the pink range in Fig. 3(a). However, when the potential V was varied from 0.01 eV to 0.15 eV, no indirect band gap was observed between band-N$_{occ}$ and band-N$_{occ}$ - 10 for the VHSs below the charge neutrality point. Here $N_{occ}$ refers to the index of
the highest occupied band in a supercell, taking into account the spin. This lack of an indirect band gap may be attributed to the competition between the detailed shape of the electronic band structure and the strength of the modulation potential.     
      From Fig. 3(a), one can see that
	 there are two global band gaps in the energy window from $\sim$E$_F$ - 150 meV to $\sim$E$_F$ + 150 meV.
	 One is located at the charge neutrality point and the other one is generated between the 
	 band-$N_{occ} + 4$ and band-$N_{occ} + 6$. The band gaps at and above the charge 
	 neutrality point are around 15 and 10 meV, respectively, when the modulation potential V is set to 0.1 eV.

	 To check the topological charge of these two gaps, we calculated the bulk 
	 Wannier center evolutions and energy dispersions of edge states. 
	 Fig. 3(b) shows the Wannier center evolutions by modulating the chemical
	 potential at the charge neutrality point. One can observe that each pair of evolution lines exhibits a fixed partner and becomes degenerate at two time-reversal-invariant points $k_x$=0 and $\pi/a$. 
  Therefore, the Wannier center evolutions do not intersect with the reference lines, indicating that the gap at the charge neutrality point is topologically trivial. However, the Wannier center evolutions switch partners at k$_x$ = 0 and $\pi$/a while keeping the occupied bands fixed at band-N$_{occ}$ + 4, as depicted in Fig. 3(c). The zigzag pattern of the Wannier center evolutions ensures an odd number of intersections with the reference lines, resulting in a non-trivial Z$_2$ band gap.
  
     The topological phases were further checked by calculating edge states with open boundary conditions along the $b$-axis. From the energy dispersion illustrated in Fig. 3(d), we are mainly interested in the three zones that are indexed by 
	 the energy windows of $\sim$E$_F$ - 0.04 eV to $\sim$E$_F$ + 0.01 eV, 
	 $\sim$E$_F$ + 0.01 eV to $\sim$E$_F$ + 0.06 eV, and $\sim$E$_F$ + 0.06 eV to $\sim$E$_F$ + 0.1 eV. In the first region, although edge bands are present within the bulk band gap, they intersect the Fermi level twice. Consequently, these edge states lack topological protection and can be eliminated through external perturbations, consistent with the behavior exhibited by the Wannier center evolutions depicted in Fig. 3(b).
    In the second region, we observe a Dirac-point-like edge band structure connecting the bulk states originating from band-N$_{occ}$ + 2 and band-N$_{occ}$ + 4, as depicted in Fig. 3(a). Notably, there exists a noticeable band anti-crossing between band-N$_{occ}$ + 2 and band-N$_{occ}$ + 4 around the X point, although there is no indirect band gap present. Therefore, the band structure resulted from band-N$_{occ}$ + 2 and band-N$_{occ}$ + 4 forms a non-trivial Z$_2$ topological semimetal state when considering the CDW modulation potential. This differs from the case in monolayer TaIrTe$_4$, where a direct non-trivial Z$_2$ band gap of approximately 5 meV is formed by band-N$_{occ}$ + 2 and band-N$_{occ}$ + 4.

	 While there is not a global band gap between band-$N_{occ}+2$ and 
	 band-$N_{occ}+4$, we do observe a direct gap formed by band-$N_{occ}+4$ 
	 and band-$N_{occ}+6$ in the range of $\sim$E$_F$ + 60 meV to $\sim$E$_F$ + 100 meV.
	 From the edge energy dispersion in this energy window, one can observe that a 
	 Dirac-point state connecting the bulk bands originating from original band-$N_{occ}+4$ 
	 and band-$N_{occ}+6$. It is a typical feature of a 2D $Z_2$ topological 
	 insulator and fully agreement with the analysis from Wannier charge evolutions 
	 in Fig. 3(c). Therefore, NbIrTe$_4$ is expected to undergo a CDW phase transition 
	 along the $b$-axis direction and results in two non-trivial $Z_2$ band gap. One is a direct 
	 band gap with a globally semimetallic nature, and the other one is an indirect band gap of around
	 10 meV. Since both gaps are not far away from the charge neutrality point, they can be achieved via weak electron doping or gating.

	\section{conclusion}
        In summary, we have predicted a dual QSHI state in monolayer NbIrTe$_4$ via first-principles calculations. In addition to the original non-trivial 
	$Z_2$ band gap at the charge neutrality point, a new global $Z_2$ band gap
	of around 10 meV can be obtained when the chemical potential is shifted to 
	the VHS point in the electron doping range. Besides, an indirect band 
	inversion in the CDW phase was found in the energy window around E$_0$ + 20 meV to 
	E$_0$ + 30 meV near the X point. Since both of these two nontrivial band gaps are 
	close to the charge neutrality point, they are expected to be experimentally detected
	after weak electron doping or gating. Moreover, the VHS point above the Fermi 
	level is mainly dominated by the $4d$ orbitals of Nb. In comparison with 
	Ta-$5d$ orbitals in TaIrTe$_4$, the dual QSHI state in NbIrTe$_4$ should 
	manifest a strong correlation effect and some more correlated topological states are expected. 

	\begin{acknowledgments}
		This work was supported by the National Key R\&D Program of China
		(Grant No. 2021YFB3501503), the National Natural Science
		Foundation of China (Grants No. 52271016 and No. 52188101),
		and Foundation from Liaoning Province (Grant No. XLYC2203080). 
		Part of the numerical calculations in this study were carried out on
		the ORISE Supercomputer.
	\end{acknowledgments}

	\appendix
	
	\bibliographystyle{ieeetr}
	\bibliography{references}
	
\end{document}